\documentclass[fleqn,twocolumn,10pt]{wlscirep}
\usepackage[utf8]{inputenc}
\usepackage[T1]{fontenc}
\usepackage{bm}

\usepackage{graphics,color,graphicx,amsmath}
\usepackage{xr}
\usepackage{times}
\usepackage{url}
\usepackage{setspace}
\usepackage{xcolor}
\usepackage{wrapfig}
\usepackage{enumitem}

\usepackage{tcolorbox}

\definecolor{lightblue}{rgb}{.68,.84,.84}
\definecolor{darkgreen}{rgb}{0.0,0.5,0.0}

\newtcolorbox{mybox}[2]{
    arc=0pt,
    boxrule=#2pt,
    colback=#1,
    width=\linewidth,
    halign=left,
}
\newtcolorbox{mywidebox}[1]{
    colbacktitle=lightblue!65!black,
    arc=0pt,
    boxrule=0pt,
    colback=lightblue,
    width=\linewidth,
    title={#1}
}

\newcounter{boxing}
\renewcommand{\theboxing}{\arabic{boxing}}

\newenvironment{boxing}[1]
{%
    \refstepcounter{boxing}%
        \begin{mywidebox}{Box \theboxing}
            \textbf{\emph{#1}\\}%
}%
{%
        \end{mywidebox}
}%

\title{Synthesis of innovation and obsolescence}

\author[1,*]{Edward D. Lee}
\author[2]{Christopher P. Kempes}
\author[2,3]{Manfred D. Laubichler}
\author[2,4,5]{Marcus J. Hamilton}
\author[6]{Jeffrey W. Lockhart}
\author[1]{Frank Neffke}
\author[2,7]{Hyejin Youn}
\author[2]{José Ignacio Arroyo}
\author[1]{Vito D.~P.~Servedio}
\author[8]{Dashun Wang}
\author[9]{Jessika Trancik}
\author[2,10]{James Evans}
\author[9]{Vicky Chuqiao Yang}
\author[11]{Veronica R.~Cappelli}
\author[1]{Ernesto Ortega}
\author[12]{Yian Yin}
\author[2]{Geoffrey B. West}
\affil[1]{Complexity Science Hub, Vienna, Austria}
\affil[2]{Santa Fe Institute, Santa Fe, USA}
\affil[3]{School of Complex Adaptive Systems, Arizona State University, Tempe, USA}
\affil[4]{Department of Anthropology, University of Texas at San Antonio, San Antonio, TX, USA}
\affil[5]{School of Data Science, University of Texas at San Antonio, San Antonio, TX, USA}
\affil[6]{University of California, Berkeley, USA}
\affil[7]{Seoul National University, Korea}
\affil[8]{Kellogg School of Management, Northwestern University, Evanston, USA}
\affil[9]{Massachusetts Institute of Technology, Cambridge MA, USA}
\affil[10]{University of Chicago, Chicago, USA}
\affil[11]{IESE Business School, Barcelona, Spain}
\affil[12]{Department of Information Science, Cornell University, Ithaca, USA}
\affil[*]{e-mail: edlee@csh.ac.at}

\begin{abstract}
Innovation and obsolescence describe the dynamics of ever-churning social and biological systems, from the development of economic markets to scientific and technological progress to biological evolution. They have been widely discussed, but in isolation, leading to fragmented modeling of their dynamics. This poses a problem for connecting and building on what we know about their shared mechanisms. Here we collectively propose a conceptual and mathematical framework to transcend field boundaries and to explore unifying theoretical frameworks and open challenges. We ring an optimistic note for weaving together disparate threads with key ideas from the wide and largely disconnected literature by focusing on the duality of innovation and obsolescence and by proposing a mathematical framework to unify the metaphors between constitutive elements.
\end{abstract}

\begin{document}
\flushbottom
\maketitle

\thispagestyle{empty}

\noindent Over the last 3.4 million years\cite{kuhnEvolutionPaleolithic2020}, we have experienced monumental changes in the quality\cite{galorPopulationTechnology2000} and pace\cite{bettencourtGrowthInnovation2007} of life due to rapid innovation and its consequences. These include technological changes like the rise of modern medicine, the transistor, the Internet, and most recently the AI revolution, besides social changes like universal suffrage and the resurgence of anti-liberalism. These have gone hand-in-hand with the demise of vacuum tubes, 1990s fashion, and, more profoundly, the destruction of biomes on an unprecedented scale\cite{vermeijNaturalHuman2011}, including the large-scale disappearance of coral reefs, melting of the cryosphere, and massive deforestation. We live in a dynamic world, one that is constantly upturned and reshaped by the inseparable forces of innovation and obsolescence. Yet, our scientific understanding of this dynamic duality is still nascent. Obsolescence, in particular, is largely overlooked compared to the celebrated and promoted narrative of innovation, creating conceptual gaps, ambiguity, and a lack of formal approaches with testable hypotheses. This theoretical imbalance is especially concerning, given several poorly understood yet increasingly apparent trends. These include a slowdown in economic productivity that underpins global growth\cite{solowContributionTheory1956, bloomAreIdeas2020}; a decline in rates of scientific and technological progress\cite{younInventionCombinatorial2015}, disruptiveness\cite{parkPapersPatents2023}, and canonical progress\cite{chuSlowedCanonical2021}; reduced innovation within the legal system\cite{leeIncreasingVolume2024}; and accelerating ecological changes to the biosphere whose consequences for biodiversity and human societies remain uncertain.

The world we inhabit lies in between the frontiers of innovation and obsolescence. 
Key to advancing the scientific understanding of the dual forces is to recognize how they are both deeply interwoven into nearly every field of inquiry\cite{hochbergInnovationEmerging2017}. Schumpeter interlinked them in the concept of ``creative destruction,'' the process of firm birth with death\cite{schumpeterTheoryEconomic1983}. Similarly, Kuhn's notion of paradigm shift centers at the conditions for innovation and obsolescence within the dynamics of science\cite{kuhnEssentialTension2000}. Darwin introduced the idea of incremental changes leading to the vast diversity of life, complemented by Kauffman's ideas about the ``adjacent possible'' to describe the set of novelties that are one step away from what is now possible\cite{kauffmanInvestigations2000}. Meanwhile, the science of science brings to the table both universal and domain-specific features, mechanisms, and structures in the process of scientific discovery\cite{fortunatoScienceScience2018, wangScienceScience2021}. The list could go on, but, unfortunately, the very breadth of the topic, the study of innovation and obsolescence, is a veritable kaleidoscope. 

Research remains fragmented, with different disciplines operating in silos, each using distinct methods and definitions. For example, minimal physics models may explain exponents for scaling relations of novelty production\cite{triaDynamicsCorrelated2015}, but how these are related to how teams generate novelty remains unclear\cite{wuLargeTeams2019}. Similarly, the connection between biological evolution and economic innovation\cite{aghionPowerCreative2021} remains elusive. In both cases, the quality and quantity of measurements hinder intellectual progress; the fossil record is sparse and incomplete\cite{mayhewBiodiversityTracks2012}, and measures for a proper ``phylogeny'' of ideas are debated\cite{zengDisruptivePapers2023}. Although innovation is undoubtedly a fundamental element of biological and social dynamics, there are seemingly insurmountable disciplinary chasms that need to be bridged.


A major difficulty is the challenge of crossing the bridge from a conceptual understanding to predictive mathematical theories. Despite foundational work clarifying conceptual elements and proposed models, the notion of what it means to innovate, the precise point at which novelties become innovations, and the multitude of different processes that one might plausibly call obsolescence leads to a flexibility in the mappings that render cumulative progress on the problem difficult (Box~1). Instead, bringing the three together---the conceptual debate, formal mathematical theories, and a diverse range of data sets---will help triangulate exactly what it means to ``innovate'' or to go ``obsolete.'' Such a combination goes beyond individual disciplines and requires extensive familiarity with various literature, mathematical fluency, and technical rigor. This calls for a new way of forging ahead by bringing together people into a coherent field of study instead of dispersing them across disciplines. 

In the face of this challenge\cite{soleEvolutionaryEcology2013, wagnerSpacesPossible2014}, we ring an optimistic note in light of a working group held at the Complexity Science Hub (CSH) in December 2022. By bringing together a range of fields under the guise of complexity science, we were able to find common conceptual ground that hopefully serves as a starting point for reorganizing our efforts, not as an ever-shifting kaleidoscope, but as elements of a shared puzzle.

\begin{figure}[t!]
\begin{boxing}{Note on conceptual ambiguity}
\label{box:ambiguity}
    There is no single, agreed-upon definition of either innovation or obsolescence. The lack of consensus underscores the need for formal, mathematical, and testable definitions, which require an iterative dialogue between ontology---defining what we aim to study---and empirical research that captures these concepts. But there are key aspects that recur. \textbf{Innovations}, in contrast to ``inventions,'' have wide impact economically\cite{schumpeterTheoryEconomic1983} or ecologically\cite{erwinConceptualFramework2021}. Naturally, this raises the question of what one means by either criterion, and for this, there is no standard quantitative definition, and there is much plastic use of the term. Innovations can also take many forms, including an improved physical form of a tool, the reuse of an existing tool for a new use, or exaptation (e.g.,~evolution of feathers from thermal regulators to flight enablers\cite{gouldExaptationMissing1982}), and the expansion of ecological niches\cite{raboskyPhylogeneticTests2017}. As a corollary, the meaning of ``one step away'' into the adjacent possible remains debated. \textbf{Obsolescence} faces an analogous quandary. Most simply put, obsolescence refers to the complement of innovation, or to have negligible impact on the dynamics of innovation and the wider ecology of the system. 
    Extinction is one avenue, but commonplace items like nails may have negligible impact on innovation, despite their omnipresence. Again, this is not a unique specification: in social systems, obsolescence, like innovation, can happen in multiple ways. It can be voluntary or involuntary, the former resulting from strategic or policy decisions to maintain limited resources and to drive innovation and productivity (we discuss this idea further in ``Fundamental variations''). Additionally, ideas or technologies can be permanently forgotten or become unpopular. Some examples may be irreversible (e.g.,~Roman cement or Damascus steel) while others are reversible\cite{wagnerSpacesPossible2014} (e.g.,~vinyl records).
\end{boxing}
\end{figure}

We bring together key concepts from various disciplines to develop a shared framework, with the five key constitutive components detailed in Box~\ref{box:fivekeys}. These are inspired by recent work that touches on two aspects that deserve more attention in the context of innovation \cite{leeIdeaEngines2024}. One is a foundational aspect that reemerges many times in the literature, but is neither fully understood nor always incorporated into models: the relationship between innovation and obsolescence, including the role of obsolescence in promoting innovation and vice versa. The other is the importance of mathematical and quantitatively testable predictions, which includes engaging with the difficulty of mapping proxies of innovation onto the dynamics thereof. By connecting these aspects in an accessible way (and skimming over some of the finer points of debate in the literature), we pose some open problems around which the emerging field can coalesce.

\begin{figure}[t!]
\begin{boxing}{Five key elements of innovation and obsolescence}
\label{box:fivekeys}
    The elements draw on ideas from across fields, demonstrating the transdisciplinary nature of the problem. We briefly summarize these here, and elaborate on them in the main text.
    \begin{enumerate}[nosep]
    	\item \textit{Space of the possible} is the set of realizable innovations, bounded by a frontier of innovation (Kauffman's ``adjacent possible''  from theoretical biology) and the boundary of obsolescence (the ``adjacent obsolescent'').
    	\item \textit{Agents} drive innovation and obsolescence, and their associated capabilities and limitations are emphasized in the social sciences.
     	\item \textit{Inventions and novelties} are precursors of widespread and successful innovations, as distinguished in economics.
    	\item \textit{Emergent constraints} are at the macro-scale, or collective processes that lie above the level of innovations and feed back into lower-level processes. Ecology is rich in such examples like niche construction.
    	\item \textit{Metric} refers to the epistemological challenge of measuring innovation and obsolescence. 
    \end{enumerate}
\end{boxing}
\end{figure}

\begin{figure*}[t]
	\centering
	\includegraphics[width=\linewidth]{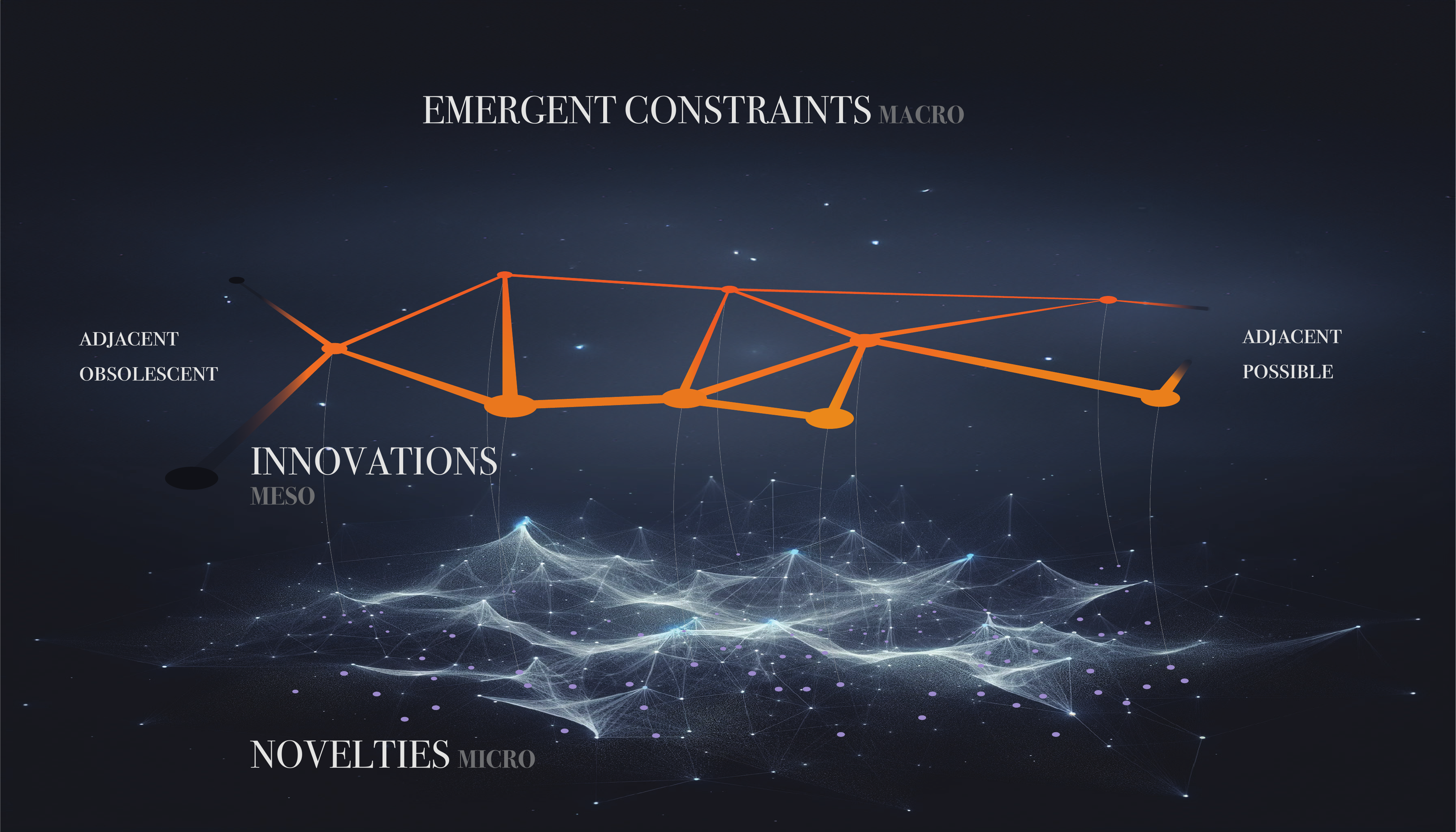}
	\caption{Conceptual diagram of the space of the possible sandwiched between the adjacent possible and the adjacent obsolescent as a 3D visualization. From the roiling set of novelties represented by the points on bottom (micro), a few rise to become innovations, the nodes on the orange graph (meso), that shape the collective environment captured by the hazy regions in the background (macro).}\label{gr:overview}
\end{figure*}


\section*{Five-part framework}
We propose a level of abstraction that is broad enough to generalize the problem across domains while remaining practical within each field. The theoretical framework boils down to five components that are highlighted in Box~\ref{box:fivekeys}: a space of the possible, agents, inventions and novelties, emergent constraints, and metrics.

The realized \textit{space of the possible} is the set of widespread technologies, mutations, or scientific theories, represented as a graph or hypergraph that connects related innovations. The space leads towards the ``adjacent possible,''\cite{kauffmanNKModel1989, kauffmanInvestigations2000} and it disintegrates at the ``adjacent obsolescent'' (the set of all things one step away from obsolescence \cite{leeIdeaEngines2024}).  
%
Naturally, the composition of the space is field-dependent. In economics, for example, the space could be manufacturing techniques\cite{mcnerneyRoleDesign2011}, product capabilities \cite{finkHowMuch2019, hidalgoProductSpace2007}, or human capabilities\cite{frankUnderstandingImpact2019, hosseiniounSkillDependencies2025}. In management science, particularly transformative innovations (e.g., ``radical,'' ``disruptive,'', or ``architectural'') are of special interest\cite{christensenWhatDisruptive2015, hendersonArchitecturalInnovation1990} because they change the nature of the market and, therewith, the competitive structure of an industry, and drive incumbents out of the market (e.g.,~digital photography, online video streaming, etc.).
Additionally, organizational innovations may involve the changing structure of the markets for technology in the 19th century\cite{lamoreauxMarketTrade2001} to the R\&D labs of the early 20th century\cite{moweryPathsInnovation1999}, and distributed innovation in multidivisional firms\cite{chandlerjrVisibleHand1993}, changing not technology but the way the firm searches for technologies. 
In biology, proposals include genotype and phenotype\cite{hosseiniExhaustiveAnalysis2015, wagnerInformationTheory2017} and, on the ecosystem level, rewiring of ecological networks\cite{huiHowInvade2019} or the emergence of symbiotic relationships that permit new metabolic pathways. In science, we often refer to the space of concepts or theories\cite{fosterTraditionInnovation2015, kwonMappingLandscape2020}. 

\textit{Agents} consist of firms, organisms, or scientists at the scale of individuals and collectives that drive the processes of innovation and obsolescence\cite{arthurFoundationsComplexity2021}.  In our framework, agents occupy the space of the possible. Some agents may be fixed to specific sites, analogous to a genotype that uniquely distinguishes groups. More generally, agents can move from one site to another, taking up or disposing of innovations flexibly, which induces a population dynamics within the space of possibilities.

\textit{Invention and novelty} are the source of innovations, or the micro-dynamics that precede fixation. These are called ``inventions'' in economics and ``genetic mutations'' in biology. This is the classic distinction between possibilities that have occurred at some point versus possibilities that have become common and widespread as are economically successful innovations,\cite{schumpeterAnalysisEconomic1935, flemingRecombinantUncertainty2001, kimTechnologicalNovelty2016, uzziAtypicalCombinations2013} fixed genes, or dominant scientific paradigms\cite{kuhnEssentialTension2000}. Sometimes inventions and novelties are distinguished from one another\cite{hochbergInnovationEmerging2017}. 

At the highest layer, we have \textit{emergent constraints}. These are shaped by the collective dynamics of agents---through cooperation, competition, or swarming---that lead to larger-scale phenomena like selection, self-organization, or social structures. These emergent dynamics create constraints that feedback into agents' actions, defining the macro-scale limits within which innovation occurs. Such limits might correspond to competitive forces in an economy in the context of firm-level innovations.  Within the context of organizational innovations within a firm, they might represent policies to shield budding inventions and allow them to mature in less constraining selection environments. In politics, media dynamics can drive polarization and regulatory stasis, such as the use of lead additives in fuel, despite evidence of its adverse impact on health\cite{mcfarlandHalfUS2022}. Along with the process of novelty generation, emergent constraints sandwich the innovation at the ``mesoscale'' as we show in Figure~\ref{gr:overview}. 

Finally, \textit{metric} refers to our limited ability to measure the processes involved in innovation and obsolescence, which limits how we can test, rule out, and infer models. For example, it is impossible to access the full set of novelties at any given time, or new production techniques may be retained as trade secrets. We must rely on lower-dimensional measures of the full process by projecting them down into simplified measures like citations, productivity, or prevalence versus performance-based metrics\cite{wuMetricsMechanisms2022}. This question is not independent of the ontology, or the definition of what we are studying, of innovation touched on above. The limits of practical choices remain to be further explored. Ideally, innovation can be studied independently of the metric, which future work may inform by connecting different metrics to one another from theoretical principles.

The key ideas are in the reduced conceptual diagram in Figure~\ref{gr:overview}. At the center of the diagram is the space of the possible, depicted as an orange graph that connects similar objects within this space. Agents possess properties, traits, or capabilities that are represented by the space, and so they live on the graph. Agent behavior drives the space to grow into the adjacent possible on the right side, which is related to the dynamics of obsolescence on the left side. This process is sandwiched between the rising foam of novelties at the bottom in purple that leads to innovations and the collective environment at the top that feeds back to constrain agent behavior, the production of novelties, and the processes of innovation and obsolescence. In any given domain, multiple coexisting processes may display such dynamics, whether they are interacting or not. Thus, the five-component framework is a minimal but multiscale representation of innovation and obsolescence, serving as a navigational map for the study of innovation and obsolescence.
%
%

As an example, we consider how scientific work on innovation can be mapped onto key elements of the diagram. Novelty creation focuses on methods of search through the adjacent possible, describing both the dynamics of the agents at the innovation front, the way they are organized, and the inherent topology of the adjacent possible\cite{triaDynamicsCorrelated2015, iacopiniInteractingDiscovery2020}. When the process of exploration is recombinatorial, the value of fruitful combinations incentivizes certain choices made in technological innovation\cite{flemingRecombinantUncertainty2001, flemingTechnologyComplex2001, younInventionCombinatorial2015}, appetizing cuisine\cite{finkSerendipityStrategy2017}, how high-performing firms collect labor skills\cite{neffkeValueComplementary2019}, etc. Sexual reproduction is a mechanism by which fruitful combinations are favored in evolution\cite{maynardsmithOriginsLife2000}. How novelties rise to become innovations touches on the evolution of multicellularity, or how complex individuals emerge that are then subject to selection forces in the environment\cite{michodTransitionsIndividuality1997, ratcliffNascentLife2017}. From another perspective, we can also examine the constraints that limit novelty exploration, whether in economic contexts like firms, or in biological contexts as favored mutations and evolutionary bottlenecks. More broadly, constraints frame scientific success\cite{acunaPredictingScientific2012} and depend on factors like individual attributes\cite{shockleyStatisticsIndividual1957, sekaraChaperoneEffect2018}, institutional prestige\cite{clausetSystematicInequality2015, wayProductivityProminence2019}, and team size\cite{wuLargeTeams2019}. Studies on the challenges of scientific innovation describe both agent population dynamics (the movement of scientists through the space of theories) and the effective constraints generated by the scientific endeavor, not least funding\cite{bushScienceEndless2021}. On the other end of the graph lies the study of obsolescence, which has received less attention. Some examples include extinction patterns\cite{newmanModelMass1997} or changes in cultural norms \cite{amatoDynamicsNorm2018}, though, as with innovation, the exact definition of obsolescence remains debated. In any given system, one might expect parallel instances of innovation-obsolescence dynamics to coexist as disconnected components. The diagram thus provides a way to contextualize together the diverse literature---far beyond the few illustrative examples that we cite here---on both innovation and obsolescence.

More formally, the elements come together in a mathematical framework coupling a living space of the possible, a temporal graph $\mathcal{G}(t)$ with vertices $x\in X$ and edges $e \in E$ changing with time $t$, which can be generalized to hypergraphs with simplicial links\cite{iacopiniSimplicialModels2019}. In addition to the changing structure, the population dynamics of agents living in the space are described by a time-dependent occupancy number $n(x,t)$\cite{leeIdeaEngines2024, leeInnovationexnovationDynamics2025}. Agent behavior determines how they push into the adjacent possible (the dynamics of the innovation front $x_i$), and obsolescence shrinks the graph (the obsolescence front $x_o$) as we show in the center of Figure~\ref{gr:overview}. The influx of novelties and effective constraints modify the parameterization of the aforementioned aspects, e.g.,~how new edges $e$ from the adjacent possible are connected to $\mathcal{G}$. In other words, the structure of the graph encodes the impact of the concomitant processes of novelty generation and environmental constraints, which in turn depend on the role of innovations. This corresponds to a set of nonlinear dynamical equations that reflects the mutual dependencies between the layers.

As a result, the diagram provides a shared language and an incipient mathematical structure for ongoing research in the field of innovation-obsolescence. In the boxes, we go through several examples discussed in the workshop that shed light on different aspects of the framework. For example, a variety of innovations in materials, manufacturing, and organization have driven steady cost reductions in the unit prices in solar cell manufacturing (Box~\ref{box:solarcell}). Another example is improvements in computational algorithms, including observations of novelty generation that fail to push the performance frontier (Box~\ref{box:algoinno}). In long-term evolutionary dynamics, novelties can be separated into distinct evolutionary phases, where innovations drive optimality given physical constraints until the constraint is modified by a ``game-changing'' innovation (Box~\ref{box:epochs}). Again, these examples indicate how the minimal formulation provides a way of incorporating previous work by bringing different aspects to light.


Thus, the framework advances three key ideas by building on the notion of a space of the possible. First, innovation and obsolescence are related. They may drive each other on---innovations rendering possibilities obsolete or obsolescence opening the door to new innovations---or they may (rarely) proceed independently. Schumpeterian creative destruction\cite{schumpeterTheoryEconomic1983} and Darwinian competition\cite{hardinCompetitiveExclusion1960} are examples of this process. 
In the former, the tension between the two forces is maximized under resource constraints\cite{weinbergerInnovationGrowth2017}, which imply zero-sum dynamics such as with van Valen's Red Queen hypothesis---although the strength of the relationship can vary. 
In many approaches, it is common to consider innovation and obsolescence as endogenous (i.e.,~innovation as the result of random mutation, and extinction as a natural outcome of competition), but they may be {\it constitutive} elements in the model, which forces an explicit accounting of the relationship between the two. Second, the evolving space of the possible is sandwiched between the microscopic dynamics generating novelty and the resulting effective constraints on the system, which then feed back into the variety of novelties explored by the system. For example, the multitude of biological mutations results in a few successful ones that can establish a new dominant ecology, which increasingly favors their survival\cite{erwinConceptualFramework2021}. This second point establishes a multiscale perspective of innovation with interaction between three scales---the micro is the froth of inventions, the meso layer is innovations, and the macro is emergent constraints (Figure~\ref{gr:overview}). Finally, the third point emphasizes the importance of developing complementary, multi-scale, and experimental perturbations to extend available quantitative metrics, which serve as limited proxies of the process. Historically, the inability to do so has mired distinctions---such as those between invention and innovation or between forgetting\cite{klinePopulationSize2010}, extinction, and obsolescence---in the conceptual realm. As is the case even with ``tangible'' physics variables\cite{changInventingTemperature2004}, inventing methods of measurement is a crucial aspect of the understanding\cite{parkPapersPatents2023}. Importantly, we emphasize that the framework is not just conceptual but also suggests a mathematical framework of stochastic, dynamical equations linking the micro, the meso\cite{leeIdeaEngines2024}, and the macro along feedforward and feedback loops.

\begin{figure}[t!]
\begin{boxing}{}
\label{box:solarcell}
    The green transition is of paramount importance today. As an example of a problem that could be mapped to the five-part framework, we discuss the falling costs of {\bf solar cell manufacturing}\cite{kavlakEvaluatingCauses2018}. One mapping of the problem to the space of the possible is as a graph of the set of combinations of cost-saving measures, where each vertex would correspond to a particular manufacturing process, and the edges connect proximate techniques. The high-dimensional space is usually projected down to cost efficiency as in Figure~\ref{gr:examples}a. The importance of cost efficiency suggests that obsolescence is primarily dictated by market forces as manufacturers eliminate uncompetitive techniques. This may describe how previous combinations of manufacturing techniques are replaced by improved ones. One notable aspect of the problem is that earlier improvements stem from ``low-level'' improvements in energy efficiency and manufacturing, but later ones stem from ``high-level'' mechanisms such as economies of scale, or that the nature of later innovations is different from that of earlier ones. 
\end{boxing}
\end{figure}

\begin{figure}[t!]
\begin{boxing}{}
\label{box:algoinno}
    How does performance improvement in artificial intelligence come about\cite{yinWorkProgress}? As another example, we discuss {\bf algorithmic innovations}, studied with a detailed data set of Kaggle data science competitions. Here, we observe each individual submission and its relative prediction performance on the entire history of submissions. As we show in Figure~\ref{gr:examples}b, there is a sea of algorithmic attempts that do not push the frontier of performance, but floating on top of it (black line) are the innovations that do and set the baseline for the next. Here, one formulation of the space of the possible is as the set of prediction algorithms for the task (each algorithm, for example, consisting of a set of components parameterized by independent performance parameters), which are then projected down to a scalar measure of performance. Performance is a natural metric because the last maximum determines the winning team. Agents correspond to the individual coding teams that are submitting to the competition. A particularly interesting aspect of this study is that we can distinguish between attempts that do not push the performance frontier, inventions and novelties. Innovations show punctuated dynamics, but these can be shown to arise from the combination of incremental and radical algorithmic attempts, capturing also the unseen layer of failures \cite{yinQuantifyingDynamics2019, yinWorkProgress}. In the broader context of the AI race today, we see the importance of emergent constraints in how algorithmic improvements alter the distribution of resources like private investment in first movers\cite{korinekConcentratingIntelligence2025}, how they change worker productivity and thus shape demand\cite{acemogluSimpleMacroeconomics2025}, and stimulate regulation that aims to shape pathways of innovation\cite{jiAIAlignment2025} such as GPDR.
\end{boxing}
\end{figure}

\begin{figure}[t!]
\begin{boxing}{}
\label{box:epochs}
      Metabolic processes in single-celled organisms have undergone several major {\bf evolutionary epochs}\cite{kempesGrowthMetabolic2012,judsonEnergyExpansions2017}. With the advent of new metabolic designs, such as the transition from prokaryotes to unicellular eukaryotes to metazoans, the relevant physical constraints change and another limit on growth rate can be saturated\cite{kempesGrowthMetabolic2012}. Within each epoch, the population devises innovations that allow organisms to saturate the growth rate given physical constraints. In terms of the five-part framework, we might think of innovations that improve metabolic efficiency within each phase as a faster process nested within slowly changing constraints that lead to discrete jumps in allometric scaling exponents\cite{delongShiftsMetabolic2010}. The latter are the metadynamics of innovation and obsolescence. The space of the possible consists of cellular structures, often projected down to rate measures, such as metabolism or growth. Selection is responsible for the emergence of new cellular functions and the disappearance of old ones, either in the sense of extinction or playing an insubstantial role in future evolutionary trajectories.
\end{boxing}
\end{figure}

\begin{figure*}[t]\centering
	\includegraphics[width=.8\linewidth]{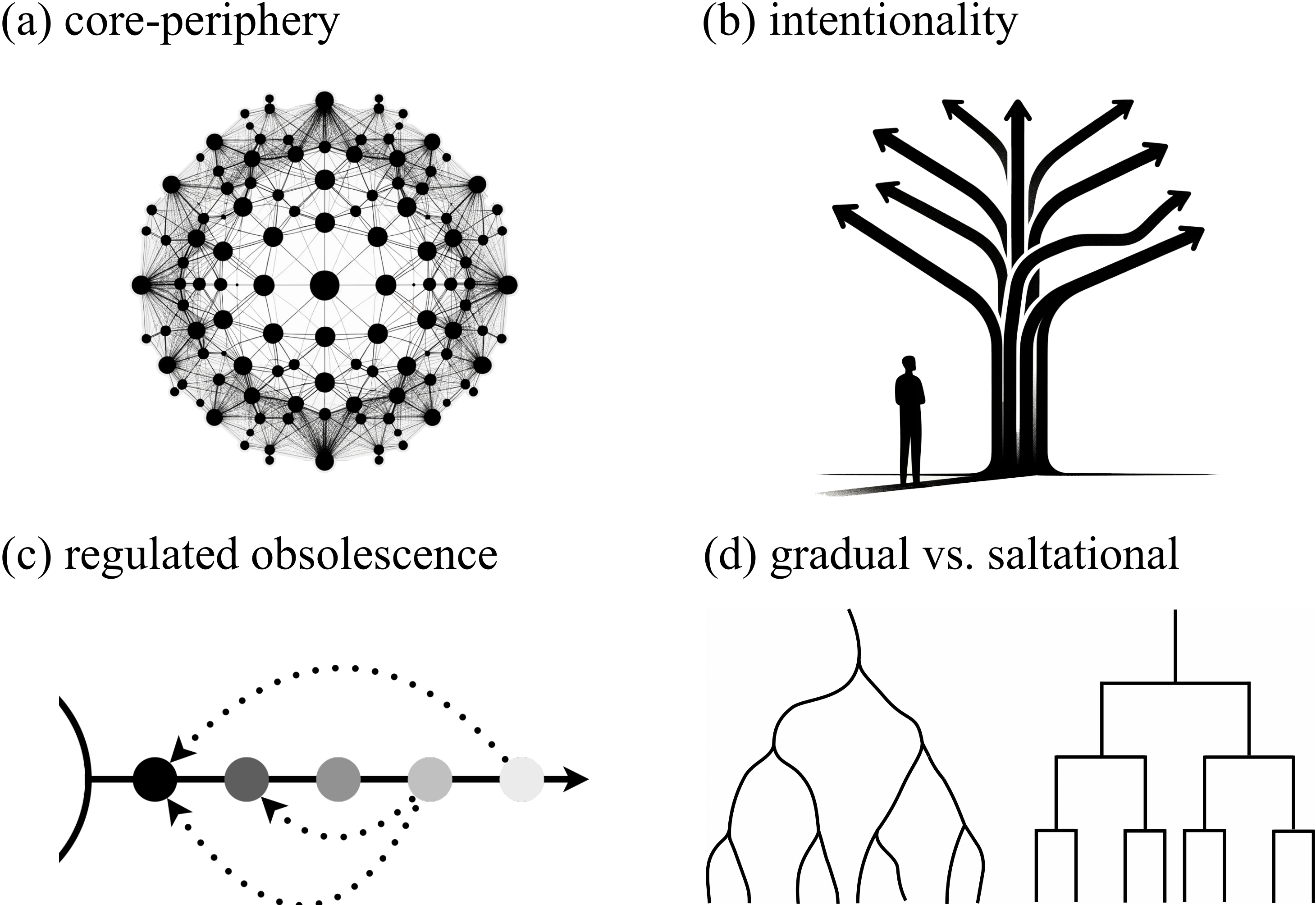}
	\caption{Diagrammatic variations on the basic framework. (a) Core-periphery. (b) Intentionality. (c) Active obsolescence. (d) Gradual vs.~saltational innovation.}\label{gr:variations}
\end{figure*}

\section*{Fundamental variations}
\begin{figure*}[t]\centering
	\includegraphics[width=.8\linewidth]{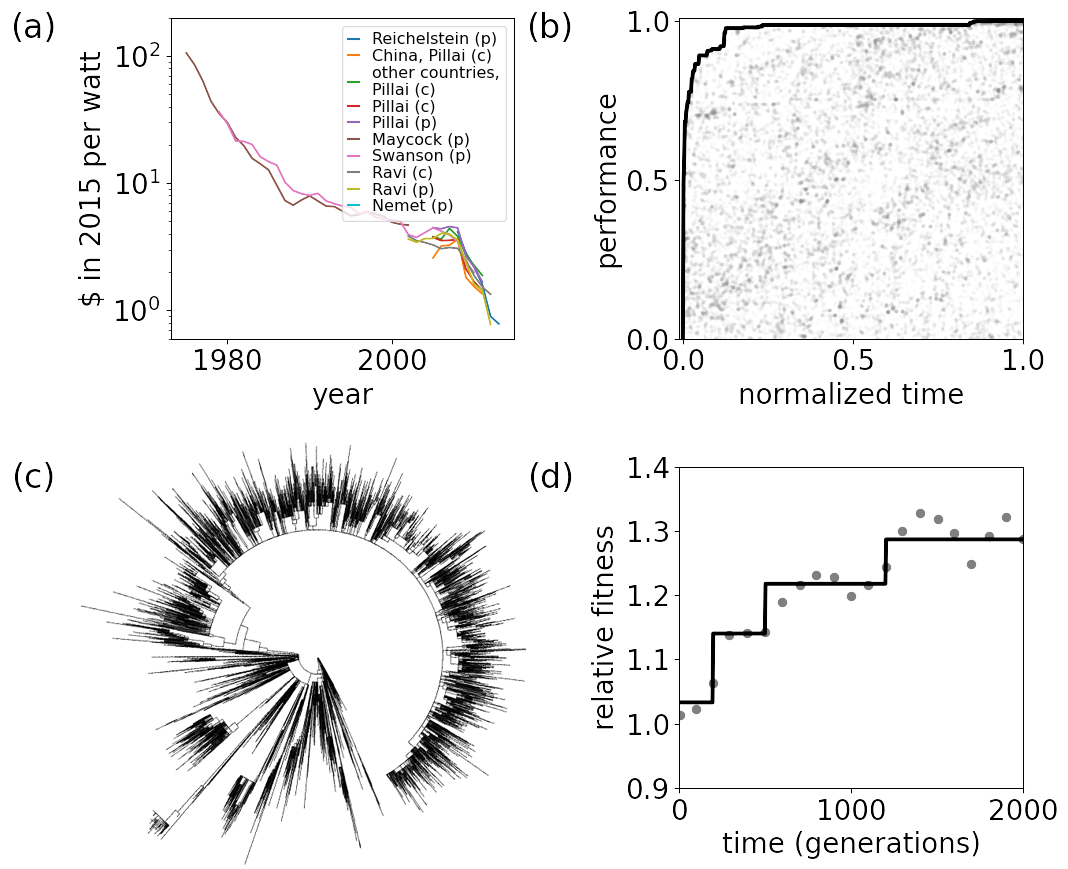}
    \caption{(a) Movement of the frontier of manufacturing innovation reflected in the falling costs of solar panels, which results from the large space of cost-cutting steps to push the productivity front from reference \citenum{kavlakEvaluatingCauses2018}. (b) Performance improvements (red line) floating on the sea of algorithmic contributions (gray points) that fail to push performance boundaries from reference \citenum{yinWorkProgress}. (c) Covid-19 phylogenetic tree for Europe, where each clade represents the set of detected innovations. Collected up until August 10, 2022. Data from reference \citenum{hadfieldNextstrainRealtime2018}. (d) An example of saltational evolution; relative fitness across generations in \textit{E.~coli} in the Long Term Evolution Experiment. Data points correspond to averages from reference \citenum{lenskiDynamicsAdaptation1994}.}\label{gr:examples}
\end{figure*}

The unified framework provides a scaffold for exploring variations across different systems, as we diagram in Figure~\ref{gr:variations}. These conceptual variations allude to some of the substantive aspects that remain untranslatable between systems. Clearly, the substrate in which innovations occur (the constituent elements of the space of the possible) and the agents that are innovating differ from system to system. The particular mechanisms by which innovations occur, innovation drives obsolescence, or vice versa, are different. Such differences do not bar statistical or structural similarities. As an example, we could group different types of relationships between innovation and obsolescence: if they happen one-to-one as in the Schumpeterian formulation, we have a case of conservation\cite{leeIdeaEngines2024}. A differential rate might be a case where innovation drives obsolescence because it displaces competitors, obsolescence drives innovation because it provides new niches for agents to thrive, or even cases of regulatory obsolescence where certain innovations are strategically removed from the space to stimulate innovation\cite{gilbertPreemptivePatenting1982}. In other words, the five-part framework refrains from committing to a particular mechanism, but lies at a level of abstraction that connects them to one another through dynamical, statistical, or functional properties, which then facilitates the comparative studies of diverse systems. 

We present several important variations to the proposed framework:

\begin{enumerate}
	\item Core-periphery of knowledge as a structural variation of the space of the possible: Innovations may be pinned to previous ideas. Scientific theories have inertia, where agents at the level of individual actors and scientific communities are organized around core concepts that are resistant to change, yet this resistance to change can itself drive innovation. A microcosm of such a phenomenon is the development of the gyroscope within the MIT Instrumentation Lab\cite{mackenzieInventingAccuracy1993}. There, a director stubbornly insisted on sticking with an older technology, a single degree of freedom floating gyroscopes, while the rest of the industry advanced. The unwillingness to question the best kind of gyroscope forced the lab to innovate in myriad peripheral areas, such as precision machining, measurement, material stability, clean room techniques, and bearing friction, in order to keep up with overall performance. The lab generated a ``protective belt'' of innovation to protect the ``hard core'' of unquestioned technology, to borrow Lakatos' terms\cite{lakatosSciencePseudoscience1977}. Refusal to innovate in one area led to compensatory innovation in others. Or, considered from a different angle, innovation in a variety of peripheral areas was used to stymie innovation in a core area. Similar refusal to abandon a core idea has led to a century of innovation in neuroscience\cite{lockhartBecauseMachine2023, lockhartLargeLong2020}. Innovation research has long thought of innovation as a directional or sequential process of recombining existing ideas in new ways, but this case highlights that the opposite matters. Human attention, knowledge, and effort are finite. Innovation requires bracketing off potential avenues. Understanding which things are overlooked and why are key to a complete understanding of innovation, including how agents see the space of the adjacent possible as a function of their relationship with the space of the possible. We picture this as a variation of the structure of the space of the possible, where some innovations (core ideas) are surrounded by a hierarchy of supporting ones like in Figure~\ref{gr:variations}a; dense local structure reduces the probability of obsolescence and even enhances the probability of innovation.
	\item Intentionality as heuristics and strategy: Human innovation follows a purposeful and biased path towards future innovations based on planning and decision making\cite{wagnerSpacesPossible2014}. Human cognition over evolutionary history has been particularly good at picking up on the available non-random mechanisms in the biosphere by specializing in things like niche construction and epigenetic-type activities, such as the intentional recombination of materials to solve adaptive solutions it identifies in the world. The ability of humans to generate variation intentionally clearly has its roots in evolved mechanisms but operates at faster scales than natural selection\cite{dennettIntentionalStance1996}. At the societal level, the purposeful generation of possibilities is manifest as \textit{exploration} centered around highly connected technologies\cite{yangCapturingValues2023}, perhaps deterred by regulatory policies\cite{sternInnovationRegulatory2017} or tied to keystone technologies\cite{bresnahanGeneralPurpose1995} that unlock new ones (e.g.,~transistors). Such anchoring may be essential\cite{kuhnEssentialTension2000}, but it also can lead to incrementalism\cite{fosterTraditionInnovation2015}. 
 More recent developments suggest that other organisms also have developed mechanisms for generating useful novelties. Such mechanisms include the potentiation of beneficial mutations\cite{wagnerArrivalFittest2014} and hypermutable states\cite{bushmanLateralDNA2002}, epigenetics that selectively modifies genetic expression\cite{gibneyEpigeneticsGene2010}, or niche construction to favor long-term strategies that can affect evolutionary trajectories\cite{lalandNicheConstruction2017}. Including these complexities into a mathematical model could build on a framework of constrained combinatorics, but perhaps more broadly as attempts at prediction and learning by heterogeneous agents, such as in game-theoretic terms, to assess the competitive landscape. 
	\item Active obsolescence and regulation as a feedback loop: In some cases, obsolescence might be a targeted or strategic process to promote innovation, such as in regulatory mechanisms. This is of particular importance, as Schumpeter noted, in a resource-limited economy, where the number of innovations that can be supported at any given time is likewise limited. Then, the investment in a new direction comes at the expense of an old one. In the free market picture, it is market forces that determine the trade-off, but the trade-off is generally also shaped by existing regulatory policies and strategic bets. In the complement, slowing obsolescence can impede innovation, such as with monopolies that attempt to stymie some types of innovation\cite{gilbertLookingMr2006} or ``patent thickets'' intended to build competitive moats or as rent-seeking\cite{gilbertPreemptivePatenting1982}. Enhanced mutation rates that move bacteria away faster from older lineages, especially directed towards loss of function, might be a form of active obsolescence in biological systems put into stressful environments\cite{bjedovStressInducedMutagenesis2003}. The strategic aspect of this trade-off deserves more attention. We might formulate this as a dynamic feedback relationship between finite resources consumed by possibilities in the adjacent obsolescent and those at the innovation front or as a game-theoretic dynamic of agents competing for resources. 
	\item Gradual vs.~saltational in terms of scales: Is the arrival of innovations gradual or punctuated? How about extinction\cite{januarioReEvaluation2021}? Again, we emphasize the importance of the dual and potentially contrasting roles of innovation and obsolescence. A long-running debate in the biological literature has been about whether evolutionary trees represent the slow accumulation of many small changes or are long periods of stasis interrupted by short periods of upheaval\cite{gouldPunctuatedEquilibria1977, vendittiSpeciationBursts2008, kondratieffLongWaves1935, freemanStructuralCrises1988, dosiTechnologicalParadigms1982}. Observed shifts in ecosystems in the fossil record or in the archaeological record of cultural evolution have not ended the debate\cite{rohdeCyclesFossil2005, valverdePunctuatedEquilibrium2015, kolodnyGameChangingInnovations2016, amatoDynamicsNorm2018}. A detailed and simultaneous view of both novelties and innovations across multiple scales would help clarify the debate around this problem.  
While the fossil record is sparse, bacteria offer a good model of study given their short generation times. The most visible example is the Long Term Evolutionary Experiment (LTEE), which propagates bacterial lines for thousands of generations. There, it has been seen that simple genotypic mutations precede and ``potentiate'' a quantum shift in the mutation rates\cite{blountGenomicAnalysis2012}, evocative of results from agent-based models\cite{christensenTangledNature2002}. This suggests one possible way of quantifying novelties (random mutations) and innovations (phenotypic shifts), but it is difficult to test models without more examples of transitions than such experiments have historically allowed (see more recent work that approaches this goal in reference \citenum{westmannEntangledAdaptive2024}). The study of innovation in other contexts may bring insight, such as punctuated progress in performance records in competitive sports\cite{arnoldRecords1998} or for algorithmic innovation. Underlying the jumps in performance are the sea of attempts at algorithmic improvement (Figure~\ref{gr:examples}b), yet we are beginning to scratch the surface of how the bed of failures matters for success\cite{yinQuantifyingDynamics2019, marianiCollectiveDynamics2024}. The deeper study of recent, detailed data sets at a fine-grained level of the innovative process will help translate the distinction into a measurable, mathematical formulation of classes to distinguish between the two paradigms originating in the study of evolution.
\end{enumerate}

\section*{New directions}
To bring these conceptual and incipient mathematical connections to life, we must first flesh out the quantitative predictions implied by the interaction between innovation and obsolescence. Second, we must take more seriously the role of mathematical predictions in explaining theory in order to build testable frameworks across the disciplines. How far do such coarse-grained theories go for specific domains? At what point do domain-specific examples require domain-specific mechanisms? In the strong form, this calls for synthesizing the threads that we have put together into a field of study, one of innovation \textit{cum} obsolescence. Such a challenge requires bringing together interdisciplinary minds to work on fundamental problems. 

One open question is how novelties become innovations, which echoes the fundamental challenge of the genotype-phenotype map\cite{hosseiniExhaustiveAnalysis2015}. This refers to the problem of determining how mutations at the level of the genetic code correspond to changes in the phenotype and the level at which selective pressure is experienced\cite{wagnerGeneDuplications2008}. An analogue in economics or science\cite{arthurNatureTechnology2009} is the patent or paper (genotype/invention), which is insufficient to know its eventual success (phenotype/innovation). A second challenge is the design of comprehensive experiments to measure the multiple scales of the problem and to validate mathematical models thereof. While the tremendous increase in data sets has led to new opportunities, studies of innovation are limited by the resolution of novelty generation and how they yield innovations, leaving untested model assumptions or parameters. Similar questions arise in computational models with artificial life\cite{maciaDistributedRobustness2009}. This calls for centers in which experimentalists and theoreticians, again across disciplines, work closer together\cite{gertnerIdeaFactory2012}. 

The challenges call for revivifying and further advancing mathematical frameworks, perhaps those that handle stochasticity in large, nonlinear dynamical systems: random matrix theory for many-body interacting and nonequilibrium systems and stochastic thermodynamics for connecting the apparent unpredictability of innovations with the question of computability\cite{fontanaArrivalFittest1994}. Existing frameworks like scaling theory that have been put to work in building the interstitial might provide the next steps, much in the way of metabolic scaling theory to compare ecosystems\cite{leeGrowthDeath2021} and as a bridge to urban ecosystems\cite{bettencourtGrowthInnovation2007, schlapferUniversalVisitation2021, hongUniversalPathway2020, younScalingUniversality2016}. 
Information theory has provided a rich set of tools and conceptual models for studying cross-disciplinary problems\cite{wagnerInformationTheory2017}. As we touch on here, networks also provide a natural and versatile way to characterize various aspects of the problem (e.g.,~how evolutionary or technological innovations interact in the space of the possible). Such tools have been forged in statistical physics\cite{soleEvolutionaryEcology2013}, representing a fertile scaffolding for extending theory. For example, certain innovations like the transistor have been transformative and thus lend themselves to the notion of phase transitions, but there are few (and fewer mathematical) connections between the two ideas in the studies of innovation or obsolescence. A shared mathematical formulation may help us start to understand how the qualitative differences in a rich conceptual area are expressed in quantitative distinctions, not least the distinction between innovative and obsolete objects. It is not only applications of existing mathematical techniques, but we imagine that bringing a rich area like innovation and obsolescence in close proximity to the quantitative realm may also inspire mathematical advances. 

For this goal, we owe an enormous debt to evolutionary biology, which has played a central and historical role in studies of innovation\cite{soleEvolutionaryEcology2013, hochbergInnovationEmerging2017, erwinConceptualFramework2021}. Not only does evolutionary biology have the longest record of roughly 3.8 billion years of innovation and obsolescence, but the field has also produced a number of formal and mathematical theories and experimental approaches addressing these challenges. In the context of evolutionary biology, innovation and obsolescence are linked through the life cycle of systems because there would be no room for novelty without death and extinction. As we discuss here, the origin of novelty is a separate problem from the success of an innovation, and the causes that explain how novelty originates are connected to the processes that govern the development and operation of these systems. These include such phenomena as recombination and the constraints that act on evolutionary space\cite{kauffmanInvestigations2000} of the possible. Finally, the success of an innovation involves complex feedback dynamics of niche construction. The effective space of possibilities depends on the specific internal and external constraints, evolving with each transformation. 
Part of our contribution is to recombine the key ideas to guide us through the broader topic of innovation and obsolescence across a large number of systems to reshape the ongoing debate on ontological and epistemological differences\cite{soleEvolutionaryEcology2013, wagnerSpacesPossible2014}.

The power of an interdisciplinary approach is that contrasting the dynamics may reveal new mechanisms that differ across the systems and allow one to test the importance of various processes by shifting them dramatically. However, it is unlikely that such problems can be genuinely addressed where the usual domain walls physically divide departments. It is a new, innovative place, free from disciplinary constraints, within or without the university walls, that will seed growth in the interstitial. The hope is that understanding the diversity of innovation dynamics across systems also leads to a new understanding of what is possible in any given system and allows us to chart the future regions of the possible into which a system might be moving. The scientific challenge is enormous, and it spans conceptual, mathematical, and empirical debates\cite{erwinProspectsGeneral2019}. With coalescence across disciplinary boundaries, we are cautiously optimistic about moving towards a unified and {\it mathematical} theory of innovation and obsolescence.


\bibliography{refs}

\section*{Acknowledgements}
E.D.L.~acknowledges funding from the Austrian Science Fund under grant ESP-127. J.W.L. acknowledges funding from the James S. McDonnell Foundation.  H.Y. acknowledges the support from the Institute of Management Research at Seoul National University and the National Science Foundation (EF 2133863). F.N.~acknowledges financial support from the Austrian Research Agency (FFG), project \#873927 (ESSENCSE). I.A.~acknowledges NSF grant 2133863. We thank Gerald Silverberg for comments on the initial proposal and the Complexity Science Hub for hosting and funding the instigatory workshop.

\section*{Author contributions}
E.D.L., C.P.K., and G.B.W.~organized the workshop. E.D.L., C.P.K., and M.L.~organized the contributors and the initial contributions into a draft. Initial draft C.P.K., E.D.L., F.N., J.W.L., M.H., and V.C.; edits C.P.K., D.W., E.D.L., G.B.W., F.N., H.Y., I.A., J.L., M.H., M.L., and V.S.; comments from E.O., J.E., J.T., E.O., V.C., and V.Y.; figures by D.W., E.D.L., I.A., J.T., and Y.Y. 

\section*{Competing interests}
The authors declare no competing interests.

\end{document}